\documentclass[pre,aps,twocolumn,showpacs,preprintnumbers]{revtex4-1}

\usepackage{graphicx}
\usepackage{dcolumn}
\usepackage{bm}
\usepackage{amssymb}
\usepackage{pifont}
\usepackage{amsmath}
\usepackage{times}
\usepackage[nooneline]{subfigure}

\begin{document}
\title{Activity statistics, avalanche kinetics, and velocity correlations
in surface growth}

\author{Juan M. L{\'o}pez}\email{lopez@ifca.unican.es}
\affiliation{Instituto de F\'{\i}sica de Cantabria (IFCA), CSIC--UC, 
E-39005 Santander, Spain}

\author{Marc Pradas}\email{pradas@ecm.ub.es}
\affiliation{Departament d'Estructura i Constituents de la Mat{\`e}ria, 
Universitat de Barcelona, Avinguda Diagonal 647, E-08028 Barcelona, Spain}

\author{A. Hern{\'a}ndez-Machado}\email{aurora@ecm.ub.es}
\affiliation{Departament d'Estructura i Constituents de la Mat{\`e}ria, 
Universitat de Barcelona, Avinguda Diagonal 647, E-08028 Barcelona, Spain}

\date{\today}

\begin{abstract}
We investigate the complex spatio-temporal dynamics in avalanche driven surface
growth by means of scaling theory. We study local activity statistics, avalanche
kinetics, and temporal correlations in the global interface velocity, obtaining
different scaling relationships among the involved critical exponents depending
on how far from or close to a critical point the system is. Our scaling
arguments are very general and connect local and global magnitudes through
several scaling relationships. We expect our results to be applicable in a wide
range of systems exhibiting interface kinetic roughening driven by avalanches of
local activity, either critical or not. As an example we apply the scaling
theory to analyze avalanches and roughening of forced-flow imbibition fronts in
excellent agreement with phase-field numerical simulations.

\end{abstract}

\maketitle

\section{Introduction} 

The dynamics of rough surfaces in systems exhibiting avalanches or bursts of
activity in response to a slow external driving has attracted a great deal of
attention in recent years. Physical examples include fracture
cracks~\cite{Maaloy.etal_2006}, fluid imbibition fronts in porous
media~\cite{Rost.etal_2007,Soriano.etal_2005,Planet.etal_2009}, or the motion of
domain-walls in magnets~\cite{Sethna_2001}. By means of high resolution fast
imaging techniques it has recently been possible to characterize experimentally
avalanche dynamics of activity from either the local velocity
map~\cite{Maaloy.etal_2006} or the average velocity time
series~\cite{Planet.etal_2009,Bramwell_2009}. A fundamental problem is to
understand the role played by local fluctuations (avalanches) in the building up
of large-scale space-time interface correlations and scale-invariant roughening.

In this paper we show that {\em local} activity statistics, avalanche scaling
properties, and time correlations (in both, the surface and the {\em global}
velocity-fluctuations) are generically connected through scaling relations that
we derive by means of simple scaling arguments. Depending on the form of
probability distribution of the first-return time of activity to a given site,
we divide avalanche driven surface growth systems into two separate classes as
follows. We consider that driven systems can exhibit either {\em off-critical}
or {\em critical} (fractal) spatio-temporal activity, and the scaling relations
among the corresponding exponents are different for each class. On the one hand,
there are systems where scale-invariant interface growth emerges from
off-critical spatio-temporal patterns of activity, where avalanches have a
finite extent and the probability for the activity to return to any given place
is short tailed. On the other hand, there are growing interfaces where local
activity is critical, showing fractal patterns in space and/or time. Examples
include systems at the depinning transition as well as interface systems that
self-organize into a critical state. This leads to critical avalanches of
activity with fat tails that lead to strong fluctuations of the avalanche size,
avalanche extent, and first return time of activity. In both cases, however, we
find that the interface space-time correlations turn out to be related to the
spatio-temporal activity, leading then to a connection between local and global
quantities via scaling relationships that we obtain. Our arguments are very
general and we expect our results to be widely applicable in systems where
surface roughness fluctuations are driven by local avalanches. 

As an example of much current interest, we apply the scaling theory to the
problem of forced-flow imbibition fronts, which occurs when a viscous fluid
advances through a disordered media displacing a less viscous fluid (typically
air)~\cite{Alava.etal_2004}. In forced-flow imbibition experiments a constant
fluid injection rate is applied while the spatially averaged velocity of the
liquid-air interface $\overline{v}$ is kept constant~\cite{Planet.etal_2009}.
This system exhibits a natural characteristic length scale
$\xi_{\times}\sim(1/\overline{v})^{1/2}$~\cite{Alava.etal_2004,Dube.etal_1999,
Laurila.etal_2005, Pradas.Hernandez-Machado_2006,MarcPradas_2009} corresponding
to the typical extent of interface correlations, which can be controlled by the
injection rate. Therefore, imbibition offers an excellent example where the
theory we shall present here can be tested, since the system may be either in an
off-critical regime ($\overline{v}\gg 0$) with avalanches extending over short
length scales of order $\xi_{\times}$, or fronts can be driven into a critical
regime ($\overline{v}\to 0$) by simply tuning the fluid injection rate.
Interestingly, these regimes can be accesible to
experimentation~\cite{Planet.etal_2009}.

The paper is organized as follows. In Sec.~II we study the local aspects of the
interface dynamics in terms of activity statistics, where we give the exact
definitions of critical and off-critical activity. We also study there the
avalanche kinetics, where we present the different scaling relations among the
local quantities describing avalanches. In Sec.~III we analyze the global
dynamics of the interface  by means of the multiscaling properties of the
surface  and the mean interface velocity time-correlations. All the different
quantities defined thoughout the paper are then studied numerically in
forced-flow imbibition fronts in Sec.~IV. The final conclusions are given in
Sec.~V.

\section{Local dynamics} 
\subsection{Activity statistics}
We consider an advancing interface $h(\mathbf{x},t)$ in $d+1$ dimensions where
$\mathbf{x}$ is the substrate position and $t$ is time. A site $\mathbf{x}$ is
said to be active at a given time if it is moving. For discrete models this
corresponds to sites to be updated, $h \to h + 1$, in the next time step. For
continuous systems defining single site activity is more cumbersome but it can
be effectively done by considering a site is active whenever the local velocity
is above some fixed threshold. In order to characterize the local
spatio-temporal activity we calculate the first-return time probability density,
$\mathcal{P}_\mathrm{f}(T)$, for a site $\mathbf{x}$ to become active again
after a period of inactivity $T$. This probability describes the time duration
of intervals separating subsequent returns of activity at any given site so that
the average number of returns ${\cal N}(T)$ in a time interval of duration $T$
satisfies~\cite{Maslov.etal_1994}
\begin{equation}
{\cal N}(T) = T - {\cal N}(T) 
\int_{\delta T}^T u \, \mathcal{P}_\mathrm{f}(u) du,
\end{equation}
where the temporal resolution can be set to unity, $\delta T = 1$, without
loss of generality.

In systems far from a critical point, which we denote as off-critical systems, 
activity is expected to be described by a general scaling function:
\begin{equation}
\mathcal{P}_\mathrm{f}(T) = a(T_0) f(T/T_0),
\end{equation}
such that $f(y)\to 0$ as $y\to\infty$ and $a(T_0)$ is a normalization constant
so that $\int^\infty du \mathcal{P}_\mathrm{f}(u) = 1$ is fulfilled. The average
return time in this case is 
simply 
\begin{equation}\label{Eq:return time}
\langle T \rangle = \int_1^\infty u \, \mathcal{P}_\mathrm{f}(u) du \propto
T_0, 
\end{equation}
and thus a finite average surface velocity $\overline{v} \propto 1/\langle T
\rangle$ exists. A very common example are systems where the return of activity
is a Poison process, so that the time interval distribution is simply
exponential $f(T/T_0)\sim \exp{(-T/T_0)}$. In the case of systems exhibiting
some degree of correlation between two consecutive returns of activity to a
given site the distribution is more likely to be described by
\begin{equation}
\mathcal{P}_\mathrm{f}(T)
\sim T^{-\tau_\mathrm{f}'} \exp(-T/T_0),
\end{equation}
where $\tau_\mathrm{f}'$ represents an effective exponent characterizing the
power-law regime. This case will also be classified as off-critical here since
there still exists a finite average surface velocity with the average return
time also given by Eq.~(\ref{Eq:return time}).

In contrast, under certain circumstances, some systems may show patterns of
activity recurrence that become critical or fractal ({\it i.e.} scale-invariant)
in time in such a way that the first-return time probability density exhibits a
power-law asymptotic tail:
\begin{equation}
\mathcal{P}_\mathrm{f}(T) \sim 
T^{-\tau_\mathrm{f}}\qquad\mathrm{for}\qquad T \gg 1, 
\end{equation}
with exponent $\tau_\mathrm{f} > 1$ for the distribution to be properly
normalizable. In this case, in the long time limit we obtain that the average
number of return points is ${\cal N}(T)\sim T^{\tau_\mathrm{f} -1}$ with some
critical exponent within the range $\tau_\mathrm{f} \in (1,2]$. Note that, for
$\tau_\mathrm{f} > 2$, although the activity is still described by a power-law,
the system is actually off-critical due to the existence of a finite return time
$\langle T \rangle$, {\it i.e.}, returns become dense in time, and the interface
advances at finite velocity ($\partial_T {\cal N} \sim \mathrm{const.}$). In
contrast, for $1 < \tau_\mathrm{f} \leq 2$ returns rarely occur and activity is
said to be fractal in time. In this case, the infinite average return time 
\begin{equation}
\langle T \rangle \sim
\int_1^\infty u^{1-\tau_\mathrm{f}} du \to \infty,
\end{equation}
typical of a critical dynamics. Fractal activity occurs, for instance, 
near a pinning/depinning critical transition ($\overline{v} \to 0$) or 
in surface growth models that self-organize into the critical state.

\subsection{Avalanche kinetics} 
An avalanche is defined as a spatially connected cluster of active sites.
Avalanches are characterized by the typical size $s (\ell)$ of an event of
lateral spatial extent $\ell$. The duration and lateral extent of an event are
related, $\ell \sim t^{1/z_\mathrm{av}}$, via the avalanche dynamic exponent
$z_\mathrm{av}$. One expects the average avalanche size to scale with the
lateral extent up to the cutoff length scale (if any): $s(\ell) \sim
\ell^{d_{\mathrm{av}}}$ for $\ell \ll \xi_{\times}$, where $d_{\mathrm{av}}$ is
the avalanche dimension exponent. Conversely, one also expects to observe a
scaling relation $s(t) \sim t^{\gamma}$ between size $s$ and duration $t$ of
avalanches. Since an avalanche of lateral length $\ell$ leads to a fluctuation
of the surface height of the order of the {\em local} width $w(l)$,
$d_\mathrm{av}$ can be easily related to the {\em local} roughness exponent
$\alpha_\mathrm{loc}$. Indeed, one has $s(\ell)\sim \ell^d w(\ell) \sim \ell^d
\ell^{\alpha_\mathrm{loc}}$, and $d_\mathrm{av}= d + \alpha_\mathrm{loc}$ for a
surface in $d+1$ dimensions. Likewise, the avalanche size-duration exponent
$\gamma$ can then be obtained from the identity $s \sim t^\gamma \sim
\ell^{d+\alpha_\mathrm{loc}}$ and we find
\begin{equation}%
 \label{gamma}
\gamma = \frac{\alpha_\mathrm{loc} + d}{z_\mathrm{av}},
\end{equation}%
which connects  avalanche dynamics with surface 
roughness~\cite{Rost.etal_2007}.

In the case of off-critical activity, non-critical avalanches with a typical
spatial extent cutoff $\xi_{\times}$ are expected so that
durations are also bounded and scaling only holds for $t \ll t_\times \sim
\xi_\times^{z_\mathrm{av}}$. The avalanche size can be written in terms of
the average velocity associated with the event as $s \sim \ell^d {\cal N}(t)
\sim \ell^d v(\ell) t$, for an avalanche of lateral extent $\ell$ and duration
$t$. We assume that the typical velocity of the avalanche 
scales with the avalanche size as
$v(\ell) \sim \ell^{-\delta}$. The identity $s \sim \ell^d v(\ell) t \sim
\ell^{d+ \alpha_\mathrm{loc}}$ leads to $\ell^{z_\mathrm{av}-\delta} \sim
\ell^{\alpha_\mathrm{loc}}$  and we obtain the scaling relation
\begin{equation}%
 \label{z_av}
z_\mathrm{av} = \alpha_\mathrm{loc} + \delta,
\end{equation}%
relating surface roughness and avalanche exponents in the case of off-critical
activity (non-critical avalanches). The exponent $\delta$ in Eq.~(\ref{z_av})
gives account of possible inter-event correlations. For instance, in the case of
uncorrelated avalanches, typically, one should expect Gaussian fluctuations of
the avalanche velocity and $\langle v(\ell)v(\ell')\rangle
\sim\delta^d(\ell-\ell')$, giving $v(\ell) \sim \ell^{-d/2}$. Avalanches are
initiated in places favored given the disordered background. If this disorder is
spatially uncorrelated it follows that avalanche velocity should also be
uncorrelated and $\delta =d/2$. Correlations in the random medium can produce
non trivial correlations in the avalanche velocity and that possibility is
contemplated here. Also, in the case of non disordered systems, where movement
is driven by some external noise, the same argument applies since the noise may
be either white or correlated. We have assumed a general situation by
introducing the exponent $\delta$ that accounts for possible velocity
correlations in case they exist. 

A different result is obtained for driven surfaces exhibiting a fractal profile
of activity which leads to critical avalanches with no characteristic length or
time scales ($t_\times \to \infty$). This is an interesting case because it
includes surfaces near the pinning/depinning transition, surfaces driven into
self-organized critical states, and surfaces obeying extremal dynamics, among
other surface models in the critical state. As discussed above, avalanches in
the presence of fractal activity do not have a finite velocity since the average
number of returns of activity to a given site is ${\cal N}(t) \sim
t^{\tau_\mathrm{f}-1}$ with $\tau_\mathrm{f} < 2$. This means that the scaling
relation (\ref{z_av}) does not apply. However, the scaling relation between
avalanche size and local width, $s \sim \ell^d {\cal N}(t) \sim \ell^d w(\ell)$,
is indeed valid also in this case, but with ${\cal N}(t) \sim
t^{\tau_\mathrm{f}-1}$. Thus we find the scaling
relation~\cite{Laurson.etal_2004}
\begin{equation}%
 \label{tau_f}
\tau_\mathrm{f} = \frac{\alpha_\mathrm{loc}}{z_\mathrm{av}} + 1,
\end{equation}%
which connects the local roughness with the statistics of the first-return time
of surfaces exhibiting a fractal activity.

\section{Global dynamics}

\subsection{Multiscaling of the height-height correlations} 
Having activity in
the form of localized avalanches implies that growth is highly inhomogeneous in
space with many time scales involved, which, in turn, may produce multiscaling
of the height-height correlations~\cite{Leschhorn.Tang_1994}. As it is shown
below, this can give information about local properties through global
observables. To this end we define $\Delta h(x,t;t_0) \equiv h(x,t+t_0) -
h(x,t_0)$ and investigate the generalized $q$-height-height correlation function
in the saturated regime
\begin{equation}\label{Cq}
 C_q(t) \equiv \langle \overline{|\Delta h(x,t;t_0) - \overline{\Delta
h(x,t;t_0)}|^q}^{1/q} \rangle \sim t^{\beta_q},
\end{equation}
where the over bar is a spatial average and brackets denote average over
independent realizations. When the exponents $\beta_q$ depend on $q$ the surface
is said to exhibit multiscaling and this indicates a highly non trivial
probability distribution of the height fluctuations. We are interested in
obtaining $\beta_q$ as a function of the roughness and activity exponents.

In the limit $q \to \infty$ only the site with the maximum growth, $\Delta
h_\mathrm{max} (t) \equiv \langle \max_x \{\Delta h(x,t;t_0)\} \rangle$,
contributes and we have $C_\infty(t) \approx \Delta h_\mathrm{max} (t)$. Note
that $\Delta h_\mathrm{max}(t)$ becomes an important quantity since it
corresponds to the typical height fluctuation produced on the surface by
avalanche events in a time interval $t$. For time intervals $t$ shorter than the
average avalanche duration cutoff ($t \ll t_\times$) we have $\Delta
h_\mathrm{max}(t)
\sim t^{\beta_\infty}$, where the exponent is $\beta_\infty \equiv
\lim_{q\to\infty} \beta_q$. In contrast, in time scales longer than the typical
event duration ($t \gg t_\times$) surface fluctuations are no longer spatially
localized--- vary many avalanches occurring at different parts of the surface
overlap and contribute to the fluctuation. Therefore surface height fluctuations
must become spatially homogeneous at long time scales ($t \gg t_\times$) and the
typical growth $\Delta h_\mathrm{max}(t)$ should scale with the global surface
width $W(t) \sim t^{\alpha/z_\mathrm{av}}$, where $\alpha$ is the {\em global}
roughness exponent. These two limiting behaviors can be put as
\begin{equation}
\label{h_max}
\Delta h_\mathrm{max}(t) \sim \left\{ \begin{array}{lcl}
     t^{\beta_\infty}    & {\rm for} & t \ll t_\times\\
     t^{\alpha/z_\mathrm{av}} & {\rm for} & t \gg t_\times
\end{array}
\right.,
\end{equation}
where the crossover time $t_\times$ is the average avalanche duration. In the
case of off-critical activity $t_\times$ is finite, while it diverges for
fractal activity.

For a system of lateral extent $L$, surface fluctuations over time scales $t \ll
t_\times$ involve localized avalanche events of activity and so only a fraction
of the surface $n_\mathrm{mov} \sim (\ell/L)^d \sim (t^{1/z_\mathrm{av}}/L)^d$
is moving. By using Eq.~(\ref{h_max}), we can roughly estimate that a term of
the form $\overline{(\Delta h)^m}$ for $m>1$ scales as 
\begin{equation}
\overline{(\Delta h)^m}\sim n_\mathrm{mov}(t)
[\Delta h_\mathrm{max}(t)]^m  \sim t^{m \beta_\infty + d/z_\mathrm{av}}L^{-d},
\end{equation}
for fluctuations at time scales $t \ll t_\times$, while it becomes
\begin{equation}
\overline{(\Delta h)^m} \sim t^{m(2\alpha+d)/2z_\mathrm{av}},
\end{equation}
for $t \gg t_\times$. Now, note that in the temporal regime $t \ll
L^{z_\mathrm{av}}$ of interest here, the main contribution in Eq.\ (\ref{Cq})
comes from the term $\overline{(\Delta h)^q}^{1/q}$. Therefore we can write $C_q
(t) \sim t^{\beta_q}$, with
\begin{equation}
\label{beta_q}
\beta_q =
\left\{ \begin{array}{lcl}
    \beta_\infty + \frac{d}{q z_\mathrm{av}} & {\rm for} & t
\ll t_\times\\
    \frac{2\alpha+d}{2z_\mathrm{av}}  & {\rm for} & t \gg
t_\times
\end{array}
\right.,
\end{equation}
that gives the multiscaling exponents $\beta_q$ induced in the surface
fluctuations by the localized avalanche dynamics, valid for both off-critical
(finite $t_\times$) and fractal (unbounded $t_\times$) distributions of
activity. The exponent for the infinite moment can then be obtained by
considering that a height fluctuation at time scales $t \ll t_\times$
corresponds to an avalanche of lateral extent $\ell$ and is given by $\Delta
h_\mathrm{max} \sim s(t)/\ell^{d} \sim t^{\gamma - d/z_\mathrm{av}}$. Using
Eq.~(\ref{gamma}) we have $\beta_\infty = \alpha_\mathrm{loc}/z_\mathrm{av}$.

It is interesting to particularize these results for the case of models
exhibiting {\em extremal} 
dynamics~\cite{Sneppen_1992,Leschhorn.Tang_1994,Paczuski.etal_1996},
where the avalanche-duration exponent is $\gamma=1$ (only one site is growing at
any given time) and we have $\Delta h_\mathrm{max}(t) \sim s(t)/\ell^d \sim
t/\ell^d$ and $\beta_\infty = 1 - d/z_\mathrm{av}$ with $t_\times \to \infty$
since the model is critical, in agreement with existing results for critical
depinning models with extremal 
dynamics~\cite{Leschhorn.Tang_1994}.

The presence of multiscaling is due to the infinitely many time scales involved
in the problem that are induced by a highly inhomogeneous surface growth. An
illustrative way to show this is to note that the time scale defined by $C_1(t)
\sim t^{\beta_1}$ is associated with the avalanche size-duration exponent
$\gamma$ since, from Eq.~(\ref{beta_q}),
$\beta_1=\beta_\infty+d/z_\mathrm{av}=(\alpha_\mathrm{loc}+d)/z_\mathrm{av}
=\gamma$ [{\it cf.} Eq.~(\ref{gamma})]. In the other limit, when $q \to \infty$,
we have $\beta_\infty = \alpha_\mathrm{loc}/z_\mathrm{av}$ that corresponds to
the time scale of single-site activity. This observation is transparent for a
fractal activity distribution where $\Delta h_\mathrm{max}\sim t^{\beta_\infty}$
can be related to the average number of returns to a given site, $\Delta
h_\mathrm{max} \sim {\cal N}(t) \sim t^{\tau_\mathrm{f} - 1}$. This gives
$\tau_\mathrm{f} = \beta_\infty + 1$, which corresponds to the previously
derived scaling relation~(\ref{tau_f}). This allows us to immediately associate
the $\infty$- correlation moment (a global observable) with the distribution of
the return times of activity at any given site (a local observable). This is
useful for experiments, where single site motion, and therefore local
observables, may be  difficult to measure.

\subsection{Correlations of the global velocity fluctuations}
Recently, there has been much
interest~\cite{Rost.etal_2007,Bramwell_2009,Planet.etal_2009,
Kuntz.Sethna_2000,Laurson.etal_2005} in obtaining information about the
avalanche kinetics from the scaling behavior of the global activity/velocity
fluctuations. This is particularly important in experiments where the global
velocity time series is readily available, while it is difficult to monitor
microscopic avalanches. The scaling relations obtained in the preceding sections
can be used to give a full solution to this problem in the context of surface
growth. 

In the stationary state, $t_0 \gg L^{z_\mathrm{av}}$, we consider the
velocity-velocity correlation $\Delta v (t) = \langle [v(t+t_0) - v(t_0)]^2
\rangle$, where $v(t)$ is the average instantaneous velocity time series.
Equivalently, we can compute the velocity fluctuation spectrum $\langle |\hat
v(\omega)|^2 \rangle$, where $\hat v(\omega)$ is the Fourier component at
frequency $\omega \sim t^{-1}$. Generally, one expects $\Delta v(t) \sim t^{2H}$
where $H$ is Hurst exponent of the velocity time series, while in Fourier space
we have $\langle |\hat v(\omega)|^2 \rangle \sim \omega^{-\nu}$, with $\nu =
2H+1$. Since $v(t) \equiv \partial_t \overline{h}$, the velocity fluctuation
spectrum can be written as $\langle |\hat v(\omega)|^2 \rangle = \omega^2
\langle |\widehat{\overline{h}}(\omega)|^2 \rangle$. We are interested here in
the case of surfaces exhibiting kinetic surface roughening in $d+1$ dimensions
where the structure factor describing the height-height correlations scales as
\begin{equation}
\langle |\hat h(\mathbf{k},t)|^2 \rangle = k^{-(2\alpha+d)}
f(kt^{1/z_\mathrm{av}}), 
\end{equation}
where $k=|\mathbf{k}|$, and the scaling function $f(u)
\sim u^{2\alpha+d}$ for $u \to 0$. We have 
\begin{equation}
\langle
|\widehat{\overline{h}}(\omega)|^2 \rangle = \lim_{\mathbf{k} \to 0}\langle
|\hat h(\mathbf{k},\omega)|^2 \rangle \sim (1/\omega)^{(2\alpha+d)/z_\mathrm{av}
+ 1}, 
\end{equation}
leading to the scaling exponent
\begin{equation}
\label{nu}
 \nu = \frac{2\alpha + d}{z_\mathrm{av}} - 1,
\end{equation}
which relates the low-frequency behavior of the velocity-fluctuation spectrum to
the roughness properties of the surface. This formula generalizes an earlier
analytical result~\cite{Krug_1991}, $\nu = (d+4)/z - 3$, valid for the
Kardar-Parisi-Zhang equation~\cite{Kardar.etal_1986}. This scaling behavior
holds for frequencies above $\omega_0 \sim (1/L)^{z_\mathrm{av}}$, below which
the spectrum becomes flat.
\begin{figure}
\includegraphics[width=0.42\textwidth]{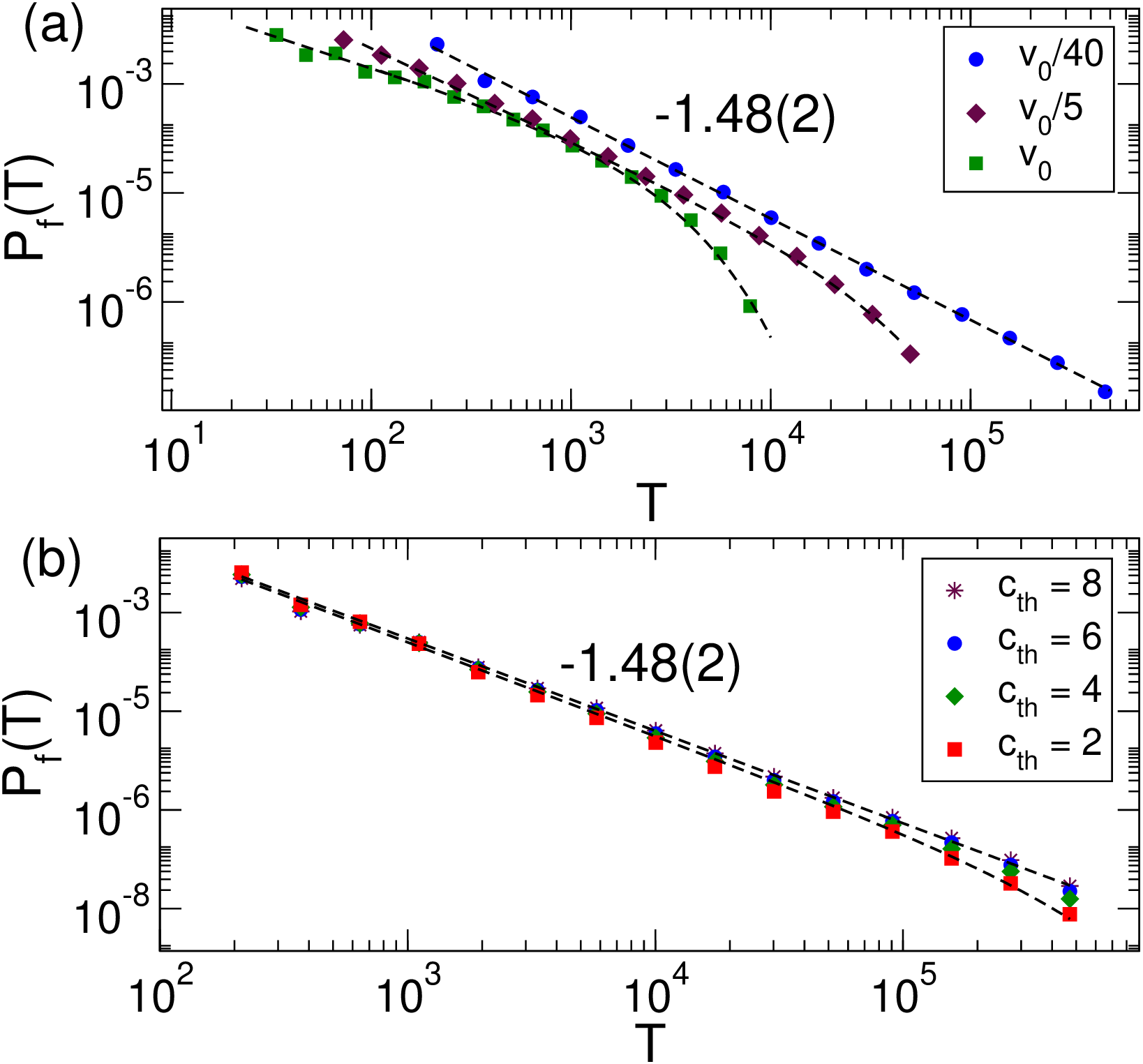}
\caption{(Color online) (a) First-return time probability density ${\cal
P}_\mathrm{f}(T)$ calculated at different velocities. It reflects an
off-critical activity (moderate velocities: $\overline{v}=v_0$ and
$\overline{v}=v_0/5$), characterized by : ${\cal P}_\mathrm{f}(T)\sim
T^{-\tau_f'}\exp{(-T/T_0)}$ with effective exponents of $\tau_f'\simeq 1.0$ and
$1.37$, respectively. Fractal activity is found at low velocities: $v=v_0/40$
where ${\cal P}_f(T)\sim T^{-\tau_f}$ with $\tau_f = 1.48 \pm 0.02$.
All results shown correspond to a threshold value of $c_\mathrm{th}=6$. (b)
Effect of the election of a velocity threshold $c_\mathrm{th}$. First-return
time probability density for the case of $\overline{v}=v_0/40$, when the
single-site activity is critical, calculated by choosing different values of the
threshold $c_\mathrm{th}$. For $c_\mathrm{th}<6$ the distribution can be fitted
to ${\cal P}_\mathrm{f}(T)\sim T^{-\tau_f}\exp{(-T/T_\mathrm{th})}$ with
$T_\mathrm{th}$ depending on $c_\mathrm{th}$. Note that $\tau_f \simeq 1.48$ is
robust for different velocity thresholds $c_\mathrm{th}$ indicating it is a
critical exponent.}
\label{fig1}
\end{figure}

\section{Case study: Fluid imbibition in random media}
We now apply the above scaling theory  to the problem of forced-flow imbibition
in disordered media~\cite{Alava.etal_2004}. As it has been noted in the
introduction, the main point in forced-flow imbibition lays in the existence of
a natural characteristic length $\xi_{\times}\sim(1/\overline{v})^{1/2}$ for the
typical avalanche
extent~\cite{Dube.etal_1999,Laurila.etal_2005,Pradas.Hernandez-Machado_2006,
MarcPradas_2009,Alava.etal_2004}, which can be controlled by the liquid flow
rate. In this way, we have a system whose activity can be made progressively
fractal as the velocity is tuned from moderate to very low values. 
\begin{figure}
\includegraphics[width=0.45\textwidth]{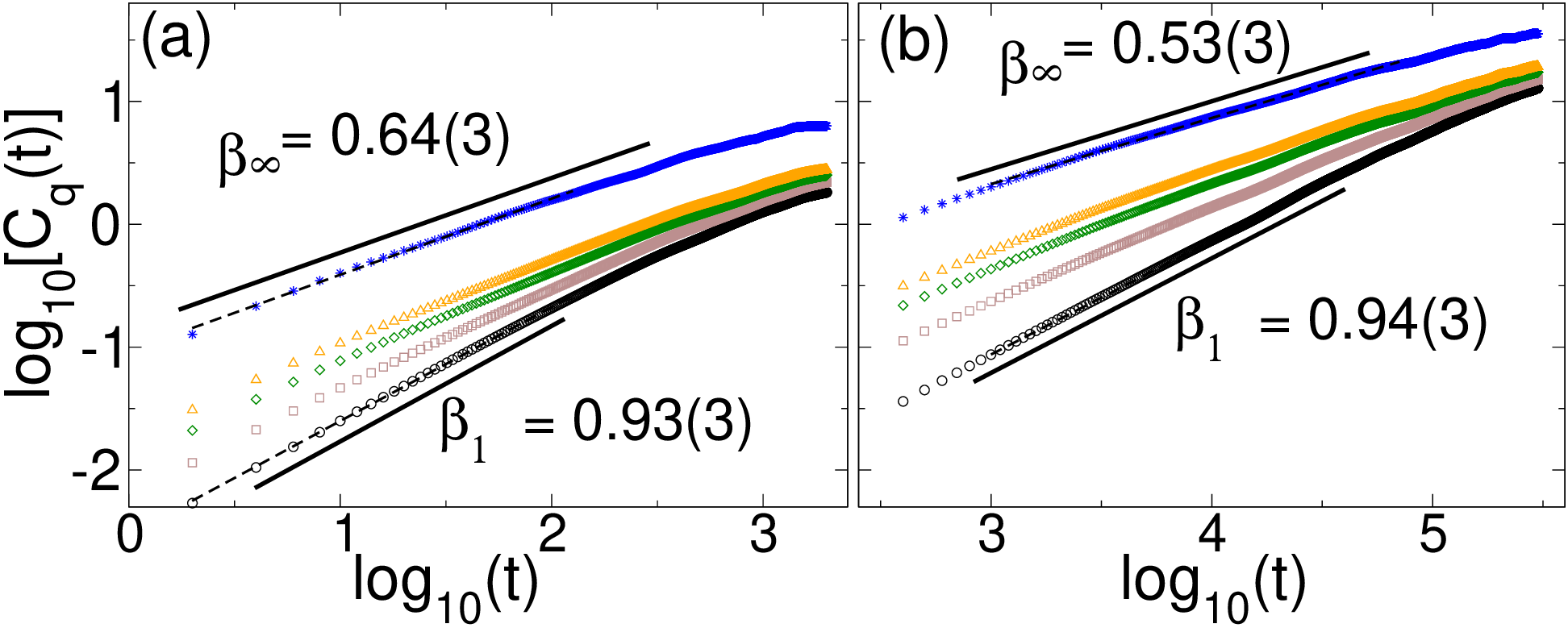}
\caption{(Color online) Height-height time correlation function $C_q(t)\sim
t^{\beta_q}$, calculated at two typical velocities, shows the presence of
multiscaling with $\beta_1 \approx 0.93$, $\beta_2\approx 0.81$, $\beta_3\approx
0.73$, $\beta_4\approx 0.70$, and $\beta_\infty\approx 0.64$ at moderate
velocities, $\overline{v}=v_0$ (a), and $\beta_1\approx 0.94$, $\beta_2\approx
0.75$, $\beta_3=0.67$, $\beta_4\approx 0.63$, and $\beta_\infty\approx 0.53$ at
low velocities, $\overline{v}=v_0/40$ (b).}
\label{fig2}
\end{figure}

We simulate forced-imbibition in $1+1$ dimensions in the capillary dominated
regime by integrating a standard phase-field model~\cite{Dube.etal_1999,
Laurila.etal_2005,Pradas.Hernandez-Machado_2006,MarcPradas_2009,Alava.etal_2004}
\begin{align}\label{eq:Phase-Field}
\partial_t\phi=\bm{\nabla}
M\bm{\nabla}\big[-\phi+\phi^{3}-\epsilon^{2}\bm\nabla^{2}\phi-\eta(\bm{r})\big],
\end{align}%
in a two-dimensional system of lateral size $L=512$ with $\epsilon=1$. The 
quenched random field $\eta(\bm{r})>0$ models capillary disorder and favors the 
liquid (wet) phase, forcing the interface to advance at the expense of the air 
(dry) phase. The parameter  $M$ in the above equation is the mobility 
 which we take constant at the liquid phase ($\phi>0$)
and zero at the air phase ($\phi<0$).
In our numerical model we have used a spatially
distributed dichotomic quenched noise and the results have been averaged over 
$25$ disorder realizations. The interface
position $h(x,t)$ separating the wet and dry phases is computed. A site on the
interface is active at time $t$ if its local velocity is above some fixed
threshold, $v(x,t) > c_\mathrm{th}\overline{v}$, where $c_\mathrm{th}$ is some
arbitrary constant. 

In Fig~\ref{fig1}(a) we plot the distribution of first-return times of activity
to any given site. We see that for moderate velocities, $\overline{v} =
v_0=2\times 10^{-3}$ and $\overline{v} = v_0/5$, the activity is an
exponentially cut-off power-law with an exponent that depends on the mean
velocity, while it tends to be  pure power-law distributed for very low
velocities, $\overline{v} = v_0/40$, where we obtain $\tau_\mathrm{f} = 1.48 \pm
0.02$ over a range of more than three decades. These distributions of activity
correspond to the fact that the characteristic avalanche size (and duration)
diverges as one gets closer to the pinned state, $\overline{v} \to
0$~\cite{MarcPradas_2009}. It is worth to mention here that the observed results
do not depend on the choice of the arbitrary threshold $c_\mathrm{th}$.
Figure~\ref{fig1}(b) shows that changing the value of $c_\mathrm{th}$ only
introduces an artificial cut-off but does not change the exponent, something
already observed in Ref.~\cite{MarcPradas_2009}.

Let us now study these two different dynamical regimes. In the case of moderate
velocities (off-critical activity) the stationary surface is
super-rough~\cite{Ramasco.etal_2000} and recent numerical work
shows~\cite{Rost.etal_2007,MarcPradas_2009} that the global and local roughness
exponents are, $\alpha \approx 5/4$ and $\alpha_\mathrm{loc} = 1$, respectively.
Avalanches in imbibition are driven by the capillary
disorder~\cite{Rost.etal_2007}, which is a random quenched and uncorrelated
field, so we should expect the Gaussian value $\delta=1/2$. Inserting these
values in Eqs.~(\ref{gamma}) and (\ref{z_av}), we find the dynamic exponent
$z_\mathrm{av} = 3/2$ and the avalanche duration exponent $\gamma = 4/3$ in
$d=1$, in excellent agreement with the existing numerical
estimates~\cite{Rost.etal_2007,MarcPradas_2009} for imbibition fronts in this
range of velocities. Figure~\ref{fig2} shows that the surface fluctuations
display multiscaling up to a given characteristic time scale. In particular, we
predict $\beta_\infty = \alpha_\mathrm{loc}/z_\mathrm{av} = 2/3$, which is in
excellent agreement with the numerical estimate in Fig.~\ref{fig2}(a). However,
Eq.~(\ref{beta_q}) with $\alpha_\mathrm{loc} =1$ and $z_\mathrm{av} = 3/2$ only
reproduces the trend of the numerical estimated exponents but not the exact
values, as a clear consequence of the strong crossover effects due to the finite
cut-off time scale $t_\times$. We expect our prediction for $\beta_q$ to become
better as $t_\times$ becomes larger (see below). 
\begin{figure}[t]%
\includegraphics[width=0.42\textwidth]{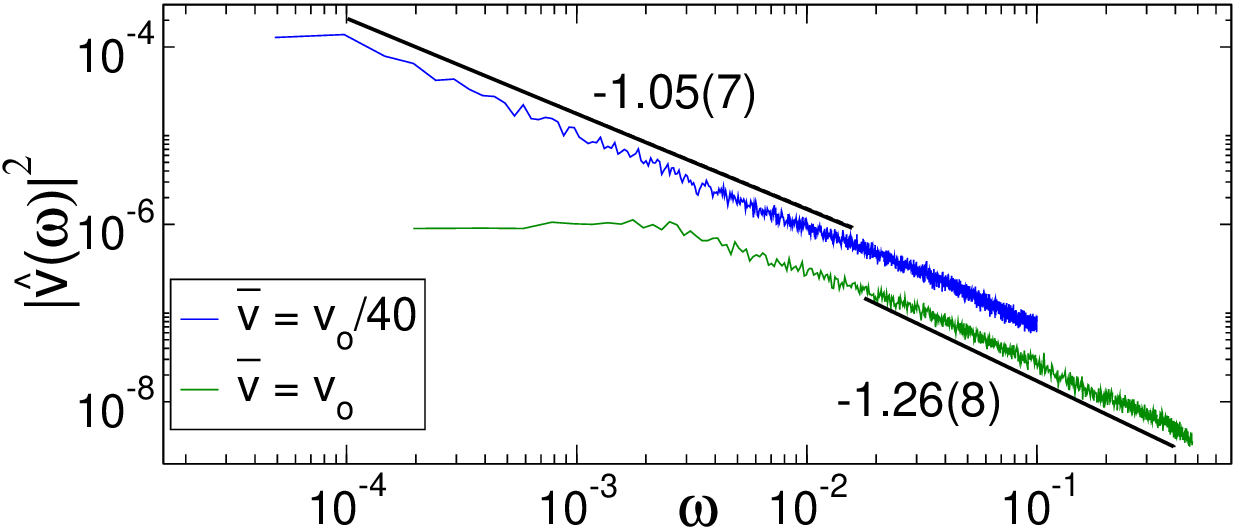}
\caption{(Color online) Power spectrum of the mean interface velocity $\langle
|\hat v(\omega)|^2 \rangle \sim \omega^{-\nu}$ calculated at  two typical
velocities, obtaining $\nu=1.26(8)$ at high velocities and $\nu=1.05(7)$ at low
velocities in agreement with the theoretical prediction $1.33$ and $1.0$
respectively.}
\label{fig3}%
\end{figure}%

Regarding the velocity fluctuations spectrum we predict a decay as $\sim
(1/\omega)^\nu$ with $\nu$ given by Eq.~(\ref{nu}) up to a cut-off frequency
$\omega_0 \sim (1/\xi_\times)^{z_\mathrm{av}}$ associated with the saturation
time in a finite-length correlated system. Replacing the known numerical
estimates~\cite{Rost.etal_2007,MarcPradas_2009} for the critical exponents
$\alpha=5/4$ and $z_\mathrm{av}=3/2$, we predict a velocity correlation exponent
$\nu = 4/3$ in good agreement with our simulations $\nu = 1.26\pm0.08$ in
Fig.~\ref{fig3}, as well as the numerical estimate in
Ref.~\cite{Rost.etal_2007}. 

The scaling behavior changes dramatically when we drive the front at very low
velocities, where the activity statistics becomes approximatively fractal (see
Fig.~\ref{fig1}). Although, for an infinite size system, a truly fractal
activity distribution would only appear in the limit $\overline{v} \to 0$, we
observe that for velocities as low as $\overline{v} = v_0/40$ our finite system
has a robust scale-invariant activity distribution ${\cal P}_\mathrm{f}(T)\sim
T^{-\tau_\mathrm{f}}$ over four decades in $T$. We recall that, according to the
arguments discussed in the previous sections, the scaling relation~(\ref{z_av})
is no longer valid for fractal activity and has to be replaced by~(\ref{tau_f}).
The scaling exponents $\alpha=3/2$, $\alpha_\mathrm{loc} = 1$,
$z_\mathrm{av}=2$, and $\gamma=1$ in this close-to-pinning regime have been
recently obtained~\cite{MarcPradas_2009} both analytically and numerically by 
using the phase-field model described above. We are therefore able to check
our scaling 
relations~(\ref{tau_f}),~(\ref{beta_q}), and~(\ref{nu}).  
Replacing $\alpha_\mathrm{loc} = 1$
and $z_\mathrm{av}= 2$ in (\ref{tau_f}) we find the first-return time
distribution exponent $\tau_\mathrm{f} = 3/2$, which is in excellent agreement
with our numerical estimate in Fig.~\ref{fig1}.

Figure~\ref{fig2}(b) shows that the system also exhibits multiscaling in the
regime of low velocities. In this regime the system is very close to the pinning
critical point (lacking characteristic lengths or time scales) and our
prediction for the multiscaling exponents $\beta_q$ is expected to be more
accurate. In fact, inserting $\alpha_\mathrm{loc} = 1$ and $z_\mathrm{av} = 2$
in Eq.~(\ref{beta_q}) we obtain $\beta_q = (q+1)/2q$, which leads to the set of
theoretical exponents $\beta_1 = 1$, $\beta_2 = 0.75$, $\beta_3 = 0.667$, and
$\beta_4 = 0.625$ in excellent agreement with our simulations [cf.\ Fig.\
\ref{fig2}(b)]. Also, in the limit $q \to \infty$ we expect to have
$\beta_\infty = \alpha_\mathrm{loc}/z_\mathrm{av} = 1/2$ to be compared with the
numerical result $\beta_\infty \approx 0.53$.

Finally, regarding the temporal correlations of the global velocity signal in
the low-velocity regime we can replace $\alpha = 3/2$ and $z_\mathrm{av} = 2$ in
Eq.~(\ref{nu}) to obtain $\nu = 1$ for the full range of frequencies, which is
to be compared with our numerical estimate $\nu = 1.05 \pm 0.07$ shown in
Fig.~\ref{fig3}.

\section{Conclusion}
To summarize, we have presented a scaling theory of surface roughening in
systems driven by avalanches exhibiting either off-critical activity
distribution, described by a power-law with a cut-off function, or critical
activity distribution, characterized by a pure power-law. Our results connect,
on the one hand, distributions of local activity with the scaling properties of
avalanches and surface roughness. On the other hand, temporal correlations in
both the surface and the global velocity signal with local activity and
avalanche kinetics turn out to be also connected by means of scaling relations.
For illustration, we have applied the theory to the problem of forced-fluid
imbibition, which can show both off-critical or critical distributions of local
activity as the the front velocity is tuned toward zero, obtaining in both cases
excellent agreement between theory and numerical results. Our predictions for
the sigle-site activity statistics, avalanche distribution, roughness, and
velocity-fluctuation correlations are novel and of interest for the experiments
in this little explored region in the context of forced-imbibition near the
depinning transition. Our arguments are very general and should be of general
validity in very different contexts where surface roughness fluctuations are
driven by local avalanches of motion, characterized by having either a parallel
or extremal dynamics.

\acknowledgments 
This work is supported by the DGI (Ministerio de Educaci\'on y Ciencia,
Spain) through Grant Nos. FIS2009-12964-C05-02 and 05.


\end{document}